\documentclass[aip,amsmath,amssymb,twocolumn,linenumbers,superscriptaddress,
]{revtex4}

\usepackage{graphicx}
\usepackage{dcolumn}
\usepackage{bm}
\usepackage[utf8]{inputenc}
\usepackage[T1]{fontenc}
\usepackage{mathptmx}
\usepackage[normalem]{ulem} 	
\usepackage[usenames,dvipsnames]{color}
\usepackage{xcolor} 
\usepackage{SIunits} 

\newif\iffinal
\finaltrue

\newcommand{\del}[1]{\sloppy{\textcolor{blue}{\sout{#1}}}} 

\newcommand{\macom}[1]{{\marginpar{\textcolor{blue}{#1}}}} 

\newcommand{\red}[1]{\textcolor{red}{#1}}

\newcommand{\vp}[0]{\ensuremath{V_\mathrm{p}}}
\newcommand{\vg}[0]{\ensuremath{V_\mathrm{g}}}
\newcommand{\rhos}[0]{\ensuremath{\widetilde\varrho_\text{JJ}}}
\newcommand{\rhosj}[0]{\ensuremath{\widetilde\varrho_\text{SJ}}}
\newcommand{\rhosurf}[0]{\ensuremath{\widetilde\varrho_\text{surf}}}
\newcommand{\rhonc}[0]{\ensuremath{\widetilde\varrho_{0}}}
\newcommand{\rhoa}[0]{\ensuremath{\varrho_\text{A}}}
\newcommand{\rhoo}[0]{\ensuremath{\varrho_\text{op}}}
\newcommand{\rhoc}[0]{\ensuremath{\varrho_\text{cov}}}

\newcommand{\rhoos}[0]{\ensuremath{\varrho_\text{op}^\text{surf}}}
\newcommand{\rhocs}[0]{\ensuremath{\varrho_\text{cov}^\text{surf}}}

\iffinal
	
	\renewcommand{\del}[1]{}

	\renewcommand{\macom}[1]{}
	
	\renewcommand{\red}[1]{#1}
\else
\fi

\begin{document}

\title{Probing defect densities at the edges and inside Josephson junctions of superconducting qubits}

\author{Alexander Bilmes}
\email[]{e-Mail: alexander.bilmes@kit.edu}
\affiliation{Physikalisches Institut, Karlsruhe Institute of Technology, 76131 Karlsruhe, Germany}
\author{Serhii Volosheniuk}
\affiliation{Kavli Institute of Nanoscience, Delft University of Technology, 2628 CJ Delft, Netherlands}
\author{Alexey V. Ustinov}
\affiliation{Physikalisches Institut, Karlsruhe Institute of Technology, 76131 Karlsruhe, Germany}
\affiliation{Russian Quantum Center, Skolkovo, Moscow 143025, Russia}
\affiliation{National University of Science and Technology "MISIS", 119049 Moscow, Russia}
\author{J\"urgen Lisenfeld}
\affiliation{Physikalisches Institut, Karlsruhe Institute of Technology, 76131 Karlsruhe, Germany}

\date{\today}

\begin{abstract}
	\centering\begin{minipage}{\linewidth}
		\textbf{
        Tunneling defects in disordered materials form spurious two-level systems which are a major source of decoherence for micro-fabricated quantum devices. For superconducting qubits, defects in tunnel barriers of submicrometer-sized Josephson junctions couple strongest to the qubit, which necessitates optimization of the junction fabrication to mitigate defect formation. Here, we investigate whether defects appear predominantly at the edges or \red{deep within the amorphous tunnel barrier of a junction}. For this, we compare defect densities in differently shaped Al/AlO$_x$/Al Josephson junctions that are part of a Transmon qubit. We observe that the number of detectable junction-defects is proportional to the junction area, and does not \red{significantly} scale with the junction's circumference, which proposes that defects are evenly distributed inside the tunnel barrier. Moreover, we find very similar defect densities in \red{thermally grown} tunnel barriers that were \red{formed either} directly after the base electrode was deposited, or in a separate deposition step after removal of native oxide by Argon ion milling.	
		}
	\end{minipage}
\end{abstract}

\maketitle 
\setlength{\parskip}{-0.25cm}
\section{Introduction}
Microscopic tunneling defects forming parasitic two-level quantum systems (TLS)~\cite{Phillips87,Anderson:PhilMag:1972} have attracted much attention in the superconducting quantum computing community due to their detrimental influence on qubit coherence~\cite{Muller:2019,Martinis:PRL:2005,Osborn2012,Stoutimore2021,Dunsworth2020}. Defects having an electric dipole moment may resonantly absorb energy from the oscillating electric field of the qubit mode, and efficiently dissipate it into the phonon~\cite{Jaeckle72} or BCS quasiparticle bath~\cite{Bilmes17}. This gives rise to a pronounced frequency-dependence of qubit energy relaxation times $T_1$~\cite{kim2008, Barends13}, while strongly coupled defects which reside in the tunnel barrier of the Josephson junction may cause avoided level crossings in qubit spectroscopy~\cite{Martinis:PRL:2005,Lupascu:PRB:2008,palomaki2010}.
Moreover, the defect's resonance frequencies may show telegraphic switching or spectral diffusion~\cite{Schloer2019,klimov2018,Burnett2019} due to their interaction~\cite{Lisenfeld2015} with a bath of thermally activated defects, and this leads to resonance frequency fluctuations of qubits and resonators, and causes qubit dephasing~\cite{Mueller:2014,Paladino2014,Burin2015,Faoro:PRB:2015, Meissner18,DeGraaf2021}.\\

Defects were found to reside at the interfaces and amorphous surface oxides of qubit electrodes~\cite{Wang:APL:2015, Bilmes19, Bilmes20}, and they may emerge on substrates due to contaminants or processing damage~\cite{Quintana:APL:2014,dunsworth2017}.
When defects are located inside the (typically amorphous) tunnel barrier of Josephson junctions, they couple most strongly to the qubit because of the concentrated electric field. 
Since the dawn of first superconducting qubits, the number of defects per junction was dramatically reduced by minimizing the junction area and thus the volume of the amorphous tunnel barrier~\cite{Steffen2005}. Nevertheless, Josephson junctions remain a vulnerability to up-scaled quantum processors, where individual qubits may spontaneously be spoiled by strongly coupled junction-defects drifting into qubit resonance~\cite{KlimovAPS2019}. This necessitates further optimization of Josephson junctions.\\

Here, we investigate whether defects are predominantly formed at the edges of a tunnel junction or deep inside the tunnel barrier. This information shall support progress towards more coherent qubits by optimizing junction fabrication or geometry. Moreover, it can provide insights to the long-standing question about the microscopic nature of the tunneling entities. 
For example, defects at the tunnel barrier edge might be formed by various species of adsorbates~\cite{deGraaf17} due to its exposure to processing chemicals and the atmosphere, while defects due to hydrogen-saturated dangling bonds~\cite{holder2013} could emerge all over the tunnel barrier due to hydrogen diffusibility in aluminum~\cite{Meng2017}.\\

\begin{figure}[htb!]
		\includegraphics[width=\columnwidth]{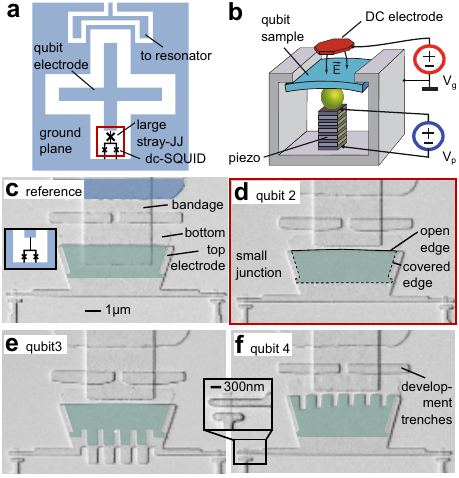}
		\caption{\textbf{Qubits with varying junction geometry.} \textbf{a} Sketch of the qubit electrode that is connected to ground via a large area ("stray") junction in series with a pair of small junctions.
		\textbf{b} Setup used to tune defects by an electric field from a voltage-biased electrode, and by mechanical strain from a piezo actuator bending the chip.
		 \red{\textbf{c - f}} Photographs of large-area junctions \red{(scale bar denotes $1\,\mathrm{\mu m}$)} with different geometries on the same qubit chip. The stray junction area is highlighted in dark cyan.
		\textbf{c} Reference qubit, where the stray junction is shorted using a bandage layer~\cite{quintana2014}. \textbf{d} Stray junction with the smallest edge length. The black solid and dashed lines indicate the junction edges that are exposed to air and covered by the top electrode metal, respectively.
		 \textbf{e} and \textbf{f}: Stray junctions with either the open or covered edge elongated by a tooth pattern. \red{The development trenches are explained in the Supplementary Methods I.} The inset in \textbf{f} shows a zoom onto a small Josephson junction. The small junction dimensions are on average $260\,\mathrm{nm}\times280\,\mathrm{nm}$.
		}
	\label{fig:1}
\end{figure}

In a previous work~\cite{Lisenfeld19}, we have shown that defects in tunnel barriers of Transmon qubits can be distinguished from those at electrode interfaces by testing their response to an applied electric field. This also revealed that qubits couple to a large number of defects residing in large-area "stray" Josephson junctions which appear as an artefact in standard shadow-evaporation~\cite{Niemeyer1976,Dolan1977,ManhattanJJ,lecocq2011} or cross-junction~\cite{Steffen2005,Wu2017} techniques. Stray junctions should thus be avoided to maximize qubit coherence, e.g. by shorting them with a so-called bandage~\cite{dunsworth2017, Bylander2020, Bilmes2021}.\\

Here, we take advantage of stray Josephson junctions for studying defects, since their larger area results in a higher number of detectable defects which improves statistics. Meanwhile, sufficient qubit coherence can be preserved since the coupling to defects in the stray junction is reduced as most of the oscillating voltage drops across the much smaller qubit junction that is connected in series. Importantly, stray junctions are formed simultaneously with the small qubit junctions and thus are expected to have identical defect densities.\\

To analyze the amount of junction-defects as a function of the Josephson junction area and the length of its perimeter, we have fabricated a series of Xmon qubits~\cite{Barends13} whose designs differ by the geometry of the large-area stray Josephson junctions. Figure~\ref{fig:1}\,\textbf{a} shows a sketch of the qubit island that is connected to ground via the stray junction in series to a small-junction dc-SQUID. The qubit electrodes were plasma-etched from a $100\,\nano\meter$-thick Al film, and the junctions were deposited with the shadow-evaporation technique after an electron-beam lithography step. Finally, Al bandages were deposited which either short the stray junction (see reference qubit in Fig.~\ref{fig:1}\,\textbf{c}) or connect it to the qubit island (see Figs.~\ref{fig:1}\,\textbf{d-f}). Fabrication details are given in Supplementary Methods I.\\

We aim to compare the concentration of defects inside the tunnel barrier with those emerging at junction edges. Moreover, we distinguish two types of junction edges: the "covered edge" that is capped by the junction's top electrode, and the "open edge" that is exposed to air (see Fig.~\ref{fig:1}\,\textbf{d}). While the area $A_\text{S}$ of each stray junction was designed to be roughly the same, the length of either the covered ($l_\text{cov}$) or open ($l_\text{op}$) edge was extended by a tooth-shape pattern as shown in Figs.~\ref{fig:1}\,\textbf{e} and \textbf{f}, respectively. Table~\ref{tab:1} summarizes the parameters of the qubits that were fabricated on two chips.\\

The standard tunneling model~\cite{Anderson:PhilMag:1972,Phillips:JLTP:1972} describes a defect by the two lowest energy eigenstates in a double-well potential, whose transition energy is $E=\sqrt{\Delta^2+\varepsilon^2}$. Here, $\Delta$ is the constant tunnel energy, and $\varepsilon=\varepsilon_\text{i}+2\textbf{pF} +2\boldsymbol{\gamma}\textbf{S}$ is the asymmetry energy given by an intrinsic offset $\varepsilon_\text{i}$ and the local strengths of electric field $\textbf{F}$ and strain $\textbf{S}$, where $\textbf{p}$ is the defect's electric dipole moment, and $\boldsymbol{\gamma}$ is its deformation potential. As described elsewhere~\cite{Lisenfeld19}, our setup (see Fig.~\ref{fig:1}\,\textbf{b}) provides in-situ control of the mechanical strain in the sample and the ability to apply a global DC-electric field, both of which can be used to tune the defect's resonance frequencies. For this work, we test the defects' response to an applied electric field to distinguish whether they are located in the tunnel barrier of a Josephson junction or at a circuit interface. Since \red{DC-electric fields inside the tunnel barrier are negligible for} a qubit in the Transmon regime~\cite{KochTransmon}, junction-defects are identified by their vanishing response to the applied E-field~\cite{Bilmes2021_npj}.\\

\section{Results}
\begin{figure}[htb]
		\includegraphics[width=\linewidth]{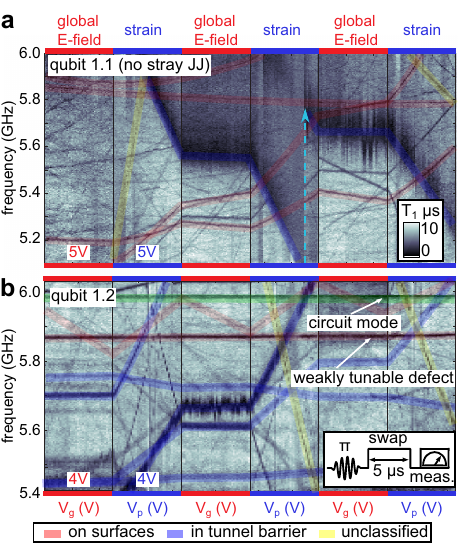}
		\caption{\textbf{Spectroscopy of defect resonance frequencies} in dependence of the applied global electric field (segments with red
		borders) and the mechanical strain (blue frames). Dark traces indicate reduced qubit $T_1$ time due to resonance with a defect. The colored shadings indicate defect locations listed in the legend. Note that in \textbf{a}, a defect in the small junction is observed, whose large coupling strength affects the qubit coherence in a wide band. Apparently, a shift of its asymmetry energy occurred during measurement (see dashed arrow), which illustrates how defects can spontaneously (dis)appear in  the qubit's spectrum. \red{The strain-tunability statistics of defects detected in Josephson junctions are analyzed in another work~\cite{Yu2021}.}
		}
	\label{fig:2}
\end{figure}

\noindent\textbf{\red{Data acquisition}}\\
To detect defects, we employ the swap-spectroscopy protocol~\cite{Cooper04,Shalibo:PRL:2010,Lisenfeld2015} for a rapid estimate of the qubit's energy relaxation time $T_1$ in dependence of frequency. In such data, Lorentzian dips in $T_1$ reveal the frequencies at which defects are resonant with the qubit~\cite{Barends13}. We repeat such measurements for a range of applied electric field and mechanical strain and alternate between both tuning channels to characterize the responses of each visible defect. Here, strain-tuning is used to sort out eventual parasitic circuit modes which show no strain response, and to increase the number of detectable defects. Figure~\ref{fig:2} shows extracts of resulting data sets, where some exemplary traces of junction-defects are highlighted in blue color, while red color marks traces of field-tunable defects residing on electrode interfaces. The yellow traces indicate non-classified defects whose location could not be identified since they were observed only during strain sweeps (blue framed data segments).\\

We then obtain a measure for the spectral density of detected junction-defects $\rhos$ by normalizing the average number of observed junction-defect traces at each applied strain to the investigated frequency range which is typically 1 GHz (see Supplementary Methods II for further details). Table~\ref{tab:1} summarizes the extracted $\rhos$ values and that of \red{non-junction defects $\rhosurf$ ("surface-defects")} detected on two sample chips.\\

The so-called shadow-junctions on chip 1 were formed using the shadow evaporation Dolan technique~\cite{Dolan1977}, where the bottom electrode is deposited, oxidized, and capped by the top electrode without removing the chip from the deposition chamber. The junctions on chip 2 were formed using the same type resist mask and identical design, however the junction's bottom electrode was exposed to air intentionally, which required Argon-milling~\cite{Gruenhaupt2017} to remove the native oxide before tunnel barrier growth. This process applied to chip 2 shall emulate so-called cross-type junctions~\cite{Steffen2005,Wu2017} whose electrodes are made in different lithography steps. In the reference qubits 1.1 and 2.1, which have no stray junctions, only few junction-defects were detected. This is an explicit verification that stray junctions increase the amount of detrimental defects coupled to the qubit, and a further affirmation that they \red{should} be omitted~\cite{Lisenfeld19}.\\
\begin{table}[htbp]
\centering
\begin{tabular}{c|c|c|c|c|c|c|c|c}
chip\&qubit & $A_\text{S}$ & $l_\text{op}$ & $l_\text{cov}$ &\rhos&\rhosurf&\rhonc& $f_{01}$ & $T_1$\\
 No.  & $\mathrm{\mu m^2}$ & $\mathrm{\mu m}$ & $\mathrm{\mu m}$ &1/GHz&1/GHz&1/GHz& $\mathrm{GHz}$ & $\mathrm{\mu s}$ \\ \hline 
 1.1 & - & - & - & 0.8 & 28.8 & 7.7 & 6.0 &  10\\
 1.2 & 12.1 & 7.1 & 10.1 & 19.6 & 10.0 & 3.5 & 6.0 & 10\\
 1.3 & 12.7 & 7.1  & 19.1 & 19.6 & 10.2 & 3.0 & 6.2 & 6\\ 
 1.4 & 14.0 & 15.7 & 11.1 & 22.5 & 24.7 & 9.7 & 6.2 &  8\\\hline
 2.1 & - & - & - & 2.3 & 67.7 & 7.1 & 5.9 &  17\\
 2.2 & 13.1 & 6.6 & 10.8 & 22.4 & 22.4 & 3.3 & 5.8 & 11\\
 2.3 & 13.6 & 6.6  & 18.4 & 23.8 & 23.8 & 3.9 & 5.7 & 12\\ 
 2.4 & 14.3 & 17.2 & 11.7 & 25.4 & 64.9 & 7.1 & 5.9 &  8\\
\end{tabular}
\caption{\textbf{Qubit parameters \red{and defect statistics}.} Chip 1 contains shadow-junctions made with Dolan bridges, while chip 2 contains cross-type junctions employing intermediate Ar ion milling. 
$A_\text{S}$ denotes the stray junction area. $l_\text{op}$ and $l_\text{cov}$ are the lengths of the open and covered stray junction edges, respectively. \rhos~ is the measured spectral density of junction-defects, \rhosurf~ that of defects on electrode surfaces, and \red{\rhonc~} that of unclassified defects. $f_{01}$ denotes the qubit's maximum resonance frequency, and $T_1$ is their average energy relaxation time. The qubit charge energy is $E_\text{C}=0.2\,\mathrm{GHz}\cdot h$, and the Josephson energies $E_\text{J}$ are $24\,\mathrm{GHz}\cdot h$ and $21\,\mathrm{GHz}\cdot h$ for samples 1 and 2, respectively. The average vacuum fluctuation strength of the qubit plasma oscillation field is $2.3\,\mathrm{kV\cdot m^{-1}}$ in the small junction, and $25\,\mathrm{V\cdot m^{-1}}$ in the stray junction. \red{In total 580 defects were detected on qubit surfaces, and 420 inside the Josephson junctions.}}
\label{tab:1}
\end{table}

\noindent\textbf{\red{Junction-defects}}\\
The defect spectral density $\rhosj$ in stray junctions is expected to be proportional to junction dimensions:
\begin{align}
\rhosj = \rhoa A_\text{S} + \rhoo l_\text{op}d + \rhoc l_\text{cov}d,\label{eq:rhoSpec}
\end{align}
where $A_\text{S}$, and $l_\text{op}$ and $l_\text{cov}$ are the respective stray junction area, and length of the open and the covered stray junction edges. The respective defect densities per GHz and per unit area are denoted by \rhoa, \rhoo~ and \rhoc.
For the effective width of the edge, we take $d\sim2\mathrm{nm}$ as discussed later. The values $\rhosj$ are obtained by subtracting from $\rhos$ (quoted in Tab.~\ref{tab:1}) the defect density inside of the small fixed-size junctions of reference qubits on the respective sample chip. The best-fitting values to Eq.~(\ref{eq:rhoSpec}) are
\begin{align}
\rhoa&=(1.5\pm0.3)\,\mathrm{(GHz\cdot\mu m^2)^{-1}}\nonumber \\ 
\rhoo&=(3\pm106)\,\mathrm{(GHz\cdot\mu m^2)^{-1}}\nonumber\\ 
\rhoc&=(62\pm114)\,\mathrm{(GHz\cdot\mu m^2)^{-1}} \label{eq:rhoFit}
\end{align}
\red{from which we deduce the relative share of defects at open and covered edges in large junctions to be on average $(0.3\pm9.1)\%$ and $(7.5\pm13.6)\%$, respectively. This, and the fact that only the fit value of \rhoa \, exceeds its fit uncertainty suggest that \rhosj~predominantly scales with the junction area, as similarly reported in previous works~\cite{Martinis:PRL:2005,Pappas2008,Osborn2012} on large-area Josephson junctions. We note that more data is required to estimate the share of junction-defects that reside at tunnel barrier edges ("edge-defects") of small junctions, as explained in more details in Supplementary Discussion I.}\\

\begin{figure}[htbp]
		\includegraphics[width=\columnwidth]{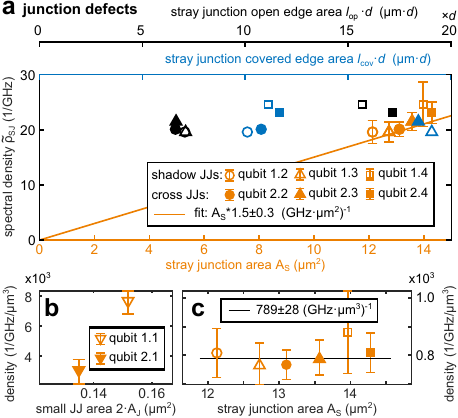}
		\caption{\textbf{Densities of defects in stray junctions} of various shapes, each embedded in a Transmon qubit circuit. \textbf{a} Spectral density $\rhosj$ of detectable defects which reside in the stray junction, plotted vs. the stray junction area (orange), or the open (black) or covered (blue) stray junction edge area (edge length times the effective edge width). The linear fit of Eq.~(\ref{eq:rhoSpec}) returns the spectral defect density per junction area inside the AlO$_x$ tunnel barrier (the slope of the orange line). \textbf{b} The density of detectable defects in the small junction of the reference qubits \red{1.1 and 1.2}, which is larger than the defect density in the stray junctions (see \textbf{c}) because of the stronger electric field. \textbf{c} Volume defect density \red{$\rhosj/(A_\text{S}d)$} which is constant for all investigated stray junctions, in agreement to the fitted slope \red{shown} in \textbf{a}.
		}
	\label{fig:3}
\end{figure}

\red{The key contribution of the junction area to \rhosj~is further} illustrated in Fig.~\ref{fig:3}\,\textbf{a} where the orange line represents the linear fit \red{reported in Eq.~\eqref{eq:rhoFit}} (see further details in Supplementary Discussion I). The error bar $\rhonc\rhosj/(\rhosj+\rhosurf)$ is the spectral density of non-classified detected defects \rhonc~multiplied with the relative part of junction-defects. Assuming a tunnel barrier thickness of $2\,\nano\meter$, the slope of the linear fit indicates a junction-defect density of $\sim760\,\,\mathrm{(GHz\cdot\mu m^3)^{-1}}$. This value is confirmed by the data in Fig.~\ref{fig:3}\,\textbf{c}, where the volume density is calculated directly from the data points of Fig.~\ref{fig:3}\,\textbf{a} and the corresponding stray junction areas. As a note, in Fig.~\ref{fig:3}\,\textbf{b} we see that the spatial density of detectable defects in the small junction is significantly larger than in the stray junctions. This is expected due to the stronger electric field inside the small junction,
which enables one to detect also defects having smaller \red{effective} dipole moments.\\

As a note, we applied the fit of Eq.~\eqref{eq:rhoFit} to merged data from both chips, since we detected the same volume density of defects (see Fig~\ref{fig:3} \textbf{c}) in both junction \red{types}. This observation indicates that dielectric losses are comparable in Josephson junctions patterned using either the shadow~\cite{Niemeyer1976,Dolan1977,ManhattanJJ,lecocq2011} or the cross-junction~\cite{Steffen2005,Wu2017} techniques.\\

\noindent\textbf{\red{Surface-defects}}\\
Detectable surface-defects are concentrated at film edges of the qubit electrodes, and it requires special methods to distinguish at which interface they reside~\cite{Bilmes20,Bilmes19}. However, the here-developed stray junction architecture allows one to independently assess densities of defects which reside at the substrate-metal and metal-air interface along the covered and open junction edges, respectively. A fit of $\rhosurf$ data quoted in Tab.~\ref{tab:1} to a linear function $\rhoos l_\text{op} + \rhocs l_\text{cov} + const$ suggests that, compared to the metal-air interface at the junction's open edge, fewer defects reside at the substrate-metal interface along the covered junction edge (see Supplementary Discussion II for more details), which is in accordance with our previous findings~\cite{Bilmes20}. Here, $\rhoos$ and $\rhocs$ are respective defect densities along the open and covered junction edge, and the offset is due to surface-defects on qubit electrodes.\\

\begin{figure}[htbp]
		\includegraphics[width=\columnwidth]{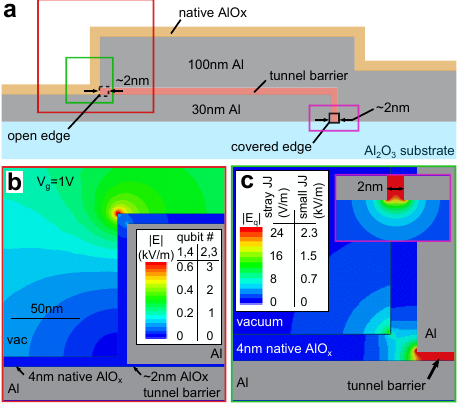}
		\caption{\textbf{Electric field distribution} near the edges of a Josephson junction, obtained with finite element simulations. \textbf{a} Sketch of the junction's cross-section (not to scale). \textbf{b} Electric field strength (color-coded) generated by the DC-gate electrode near the open edge of the junction. The legend distinguishes between qubits due to their different positions relative to the gate electrode.
		We see that the open edge is screened by the junction's top electrode so that defects residing there are not field-tunable and cannot be confused with surface-defects.
	 \textbf{c} Strength of the electric field induced by the qubit plasma oscillations at the open and covered (see inset) edges. The color bar indicates field strengths in the small and stray junctions.
	}
	\label{fig:4}
\end{figure}

\noindent\textbf{\red{Discussion of defect locations}}\\
The distribution of DC-electric fields generated by the DC-gate was simulated using the ANSYS finite element solver, to test whether edge-defects are exposed to the applied E-field. Figure~\ref{fig:4}\,\textbf{a} contains a simplified sketch of a Josephson junction's profile, where the black continuous and black dashed squares indicate the cross-sections of the covered and open junction edges, respectively. The region emphasized with a red square is magnified in Fig.~\ref{fig:4}\,\textbf{b}, showing the field distribution at the open junction edge when $1\,\mathrm{V}$ is applied to the DC-gate. The DC-potential of the top and bottom junction electrodes were set to zero as it is the case for Xmon (i.e. grounded Transmon) qubits.\\

As visible in Fig.~\ref{fig:4}\textbf{b}, the $\sim10\,\mathrm{nm}$ wide region around the open edge is free of applied DC fields due to screening by the junction's top electrode. For the same reason, the applied DC fields are zero at the covered edge. We thus can be sure that edge-defects are not field-tunable, and cannot be confused with surface-defects. Note that each qubit on the same chip couples differently to the DC-gate electrode, which is captured in the legend of Fig.~\ref{fig:4}\,\textbf{b}.\\

The AC-electric field strength induced by the qubit plasma oscillation at the open junction edge is shown in Fig.~\ref{fig:4}\,\textbf{c} where the inset shows the qubit field at the covered edge, and the legend indicates the field strengths in the small and the stray junctions. One recognizes that the qubit field is strongly confined in the tunnel barrier, and decays very fast outside, on a length scale of the tunnel barrier thickness $d\sim2\,\mathrm{nm}$. This means that surface-defects which reside within a distance range of ca. $2 - 10\,\mathrm{nm}$ to the open edges of the stray junctions (where they are not field-tunable), as well as defects which reside at the substrate-metal interface close to the stray junction's covered edge, don't couple to the qubit, and cannot be confused with edge-defects.
This is not necessarily the case at the small junctions' edges where the qubit fields are significantly larger. However, the contribution of the small junctions to the measured junction-defect density is only a small offset as mentioned before.\\

\section{Discussion}
We have studied densities of microscopic material defects in Josephson junctions of various shapes using superconducting Transmon qubits. We observed that in \red{$\sim13\,\mathrm{\mu m^2}$ large}  Al/AlO$_x$/Al Josephson junctions fabricated using shadow evaporation and thermal tunnel barrier growth, the total amount of detectable junction-defects does not \red{significantly} scale with the junction edge lengths which were varied by a factor of two, but with the junction area.\\

Thus, relevant defects seem to be evenly distributed all over the tunnel barrier of the Josephson contact, \red{which supports the old-standing strategy to reduce the dielectric losses of a Josephson junction by minimizing its footprint. We note that the size of our data set acquired on large junctions does not allow us to predict the relative share of edge defects in submicron-sized junctions.
This possibly could be investigated using the here-reported method applied to large-area and high aspect-ratio Josepshon junctions~\cite{Pop2012} where the effect of junction edges is amplified like in small junctions, while the advantage of good defect statistics is preserved.}\\

As an outlook, we \red{emphasize} that the here-presented technique to study junction-defects in large-area stray Josephson junctions is also suitable to study how the defect density and qubit coherence scale with the tunnel barrier thickness, which is another open question on the way to improved junctions.\\

We further observe that AlO$_x$ tunnel barriers, which were thermally created A in-situ after deposition of the bottom junction electrode, and B after application of Argon-milling to the bottom electrode, show the same density of detectable defects. This indicates that relevant defects are formed due to structural disorder rather than contamination from the Argon plasma, like implanted Argon ions and re-deposited mask and substrate residuals. This confirms that shadow~\cite{Niemeyer1976,Dolan1977,ManhattanJJ,lecocq2011} and cross junctions~\cite{Steffen2005,Wu2017} are equally suitable for high-coherence qubits~\cite{Schoelkopf2011,Wu2017}.\\

As a note, we observe that the density of \red{surface-defects} scales only with the open edge length of the \red{large-area} junction\red{. This} indicates that the interface of aluminum to the sapphire substrate does not notably contribute to the amount of detectable defects, which is in agreement with our previous studies~\cite{Bilmes19,Bilmes20}, and leaves room for speculations. For example, defects at the substrate-metal interface might be screened by the superconducting condensate~\cite{Bilmes17}, while at the metal-air interface defects are separated from the metal by the native Al oxide.\\

\section{Methods}

\noindent\textbf{Sample fabrication}\\
The Transmon electrodes and the readout circuitry were patterned into a $100\,\nano\meter$ thick Aluminum groundplane in an inductively coupled plasma (ICP) device, using an S1805 optical resist mask. The small and the stray Josephson junctions were simultaneously deposited in a thermal evaporation PLASSYS device using a double-resist mask patterned by eBeam-lithography. The bottom junction electrode consisted $30\,\nano\meter$ thick Al which was deposited at an angle of $50^\circ$, and at a rate of $1\,\nano\meter/\second$. For sample \#1, the $100\,\nano\meter$ thick Aluminum top electrode was deposited without breaking vacuum at the same rate but at zero tilt, after creation of the AlO$_x$ tunnel barrier (static oxidation, exposure of $1100\,\sec\cdot\milli\mathrm{Bar}$). The junctions on sample \#2 were made using the same design and type of resist mask, however  the bottom electrode was exposed to air after deposition of the bottom electrode, so that an Argon milling step was required to clean off the oxide from the bottom electrode prior thermal growth of the tunnel barrier and the successive deposition of the top electrode. After lift-off of the junction layers, a further electron-beam lithography step was applied, and aluminum bandages were placed to selectively either contact or short the stray junction. See Supplementary Methods I for further details.\\

\noindent\textbf{Supplementary Methods and Discussion}\\
In Supplementary Methods II, the detection and counting method of defects is presented and additional raw data plots like in Fig.~\ref{fig:2} are shown. Supplementary Discussion I and II contain additional analysis of defect densities in junctions and at other qubit interfaces.\\

\subsection*{Code availability}\noindent
Code is available upon reasonable request.

\subsection*{Data availability}\noindent
Data is available upon reasonable request.

\subsection*{Acknowledgements}\noindent
AB and JL gratefully acknowledge funding by Google LLC. JL is grateful for funding from the Deutsche Forschungsgemeinschaft (DFG) for project LI2446/1-2 and for funding from the Baden-W\"urttemberg-Stiftung. We acknowledge support by the KIT-Publication Fund of the Karlsruhe Institute of Technology. We acknowledge support from the German Ministry of Education and Research (BMBF) within the project GeQCoS. A.V.U. acknowledges support from the Russian Science Foundation, project No. 21-72-30026.

\subsection*{Author contributions}\noindent
The qubit samples were designed by AB, and fabricated by AB with assistance of SV. Experiments were devised and performed by JL in a setup implemented by AB and JL. AB performed E-field simulations and analyzed the data. The manuscript was written by AB and JL with contributions from all authors.\\

\subsection*{Competing interests}\noindent
The authors declare no competing interests.

\subsection*{References}\noindent
\bibliography{Biblio}

\clearpage

\setlength{\parskip}{-0.25cm}
\onecolumngrid
\renewcommand{\thetable}{\arabic{table}}
\renewcommand{\figurename}{Supplementary Figure}
\renewcommand{\tablename}{Supplementary Table}
\setcounter{figure}{0}

\begin{center}
\large
\textbf{Probing defect densities at the edges and inside Josephson junctions of superconducting qubits\\}
\vspace{0.3cm}
\textbf{Supplementary Material}
\normalsize
\end{center}

\section*{Supplementary Methods I}
\label{ss_fab}
\noindent\textbf{Sample fabrication}\\
\begin{enumerate}
\item Wafer preparation (3"):
	\begin{enumerate}
		\item Clean with piranha solution (mix of sulfuric acid H$_2$SO$_4$ and hydrogen peroxide H$_2$O$_2$) to remove organic residuals.
		\item Oxygen plasma clean in the "barrel asher" (see Tab~\ref{tab_fab}): $10\,\minute$ duration, $45\,\mathrm{sccm}$ O$_2$ at a chamber pressure of $0.7\,\milli\mathrm{Bar}$, $150\,\watt$ generator power.
		\item After the O2 clean, transfer ($\sim 5\,\minute$) the wafer into the PLASSYS shadow evaporator and pump the load lock for $2\,\hour$ at $200\,\mathrm{^\circ C}$. Cool down to room temperature over night. Final pressure is $\sim1\cdot10^{-7}\,\milli\mathrm{Bar}$.
		\item Apply oxygen-argon cleaning recipe (Table~\ref{tab_descum}) for $15\,\mathrm{sec}$ at $45^\circ$ tilt angle.
		\item Ti gettering (deposit $\sim30\,\nano\meter$ Ti into the load lock with closed shutter, and wait for $5\,\minute$).
	\item Deposit $\sim100\,\nano\meter$ Al that will define the qubit ground plane (zero tilt, deposition rate of $1\,\nano\meter/\second$).
	\item Apply static oxidation ($10\,\minute$ at $10\,\milli\mathrm{Bar}$) to passivate the Al film surface.
	\item Spin coat the wafer with S1818 protecting resist, and dice it into seven $2\,\mathrm{cm}\times2\,\mathrm{cm}$ wafers.
	\item Remove the protecting resist in a NEP bath ($1\,\hour$ at $90\,^\circ\mathrm{C}$).
	\end{enumerate}
\item Definition of qubit electrodes and readout resonator:
	\begin{enumerate}
		\item Spin coat the wafer with S1805 resist ($\sim 300\,\nano\meter$ thick, $6$k rpm, $60$ sec, 1 min on hotplate at 115 $^\circ$C), and apply positive optical lithography using the mask aligner (Tab.~\ref{tab_fab}).
		\item Etch the Al film using an Ar-Cl plasma in an ICP device (Tab.~\ref{tab_fab}).
		\item Remove the resist in a NEP bath ($1\,\hour$ at $90\,^\circ\mathrm{C}$).
	\end{enumerate}
\item Deposition of the Josephson junctions:
	\begin{enumerate}
		\item Spin coat the wafer with a double resist ($\sim250\,\nano\meter$ PMMA A-4 on top of $\sim900\,\nano\meter$ MMA EL-13, each spun at $2$k rpm for $100$ sec, and baked for $5$ min at $200\,^\circ$C), and apply positive electron-beam lithography using the JEOL device (Tab.~\ref{tab_fab}).\\
		\red{NOTE: the development trenches indicated in Fig.1 \textbf{f} enable the $\sim 2\,\mathrm{\mu m}$ deep suspension of the imaging top resist in a sufficiently short time ($\sim 100\,\mathrm{s}$ in IPA at $6\,^\circ\mathrm{C}$), to prevent overdevelopment of the Dolan bridge which defines the small Josephson junction.}
		\item Oxygen plasma clean with the barrel asher ($3\,\minute$ duration, $7.5\,\mathrm{sccm}$ O$_2$, $230\,\watt$ generator power) to remove resist residuals.
		\item Load the wafer into the PLASSYS and pump the load lock for $\sim2\,\hour$ at room temperature to a pressure of $\sim4\cdot 10^{-7}\,\milli\mathrm{Bar}$.\label{loadplassys}
		\item Apply oxygen-argon cleaning recipe (Table~\ref{tab_descum}) for $15\,\mathrm{sec}$ at zero tilt.
		\item Ti gettering (deposit $\sim30\,\nano\meter$ Ti into the load lock with closed shutter, and wait for $5\,\minute$).\label{tigett}
		\item Deposit $30\,\nano\meter$ Aluminum at $50^\circ$, and at a rate of $1\,\nano\meter/\second$, in order to define the junction's bottom electrode.
		\item {\textbf{sample \#1:}}
		\subitem- in-situ, apply static oxidation (exposure $220\,\second$ at $5\,\milli\mathrm{Bar}$) to form the AlO$_x$ tunnel barrier.
		\subitem- Deposit $100\,\nano\meter$ Aluminum at zero tilt, in order to define the junction's top electrode.
		\subitem- Lift-off in a NEP bath ($1\,\hour$ at $90\,^\circ\mathrm{C}$).
		\item {\textbf{sample \#2:}}
		\subitem- unload sample from vacuum chamber and store at air for at least 1 day.
		\subitem- apply steps \ref{loadplassys},\ref{armill} and \ref{tigett}
		\subitem- apply static oxidation (exposure $290\,\second$ at $10\,\milli\mathrm{Bar}$) to form the AlO$_x$ tunnel barrier. Note that $J_\text{C}$ of junctions on both samples \#1 and \#2 were very similar despite different exposures, due to Argon milling which changed the physical and chemical property of the bottom electrode surface of sample \#2.
		\subitem- Deposit $100\,\nano\meter$ Aluminum at zero tilt, in order to define the junction's top electrode.
		\subitem- Lift-off in a NEP bath ($1\,\hour$ at $90\,^\circ\mathrm{C}$).
	\end{enumerate}
\item Bandaging of the junctions:
	\begin{enumerate}
		\item Spin coat the wafer with a double resist (same parameters as before), and apply positive electron-beam lithography using the JEOL device (Tab.~\ref{tab_fab}).
		\item Load the wafer into the PLASSYS and pump the load lock for $\sim2\,\hour$ at room temperature to a pressure of $\sim4\cdot 10^{-7}\,\milli\mathrm{Bar}$.
		\item Apply argon milling (Table~\ref{tab_armill}) for $2.5\,\minute$ at zero tilt, in order to remove the native oxide. \label{armill}
		\item Ti gettering (deposit $\sim30\,\nano\meter$ Ti into the load lock with closed shutter, and wait for $5\,\minute$).
		\item Deposit $150\,\nano\meter$ Aluminum at zero tilt, in order to define the bandage which either connects the stray junction (qubits 2-3), or shorts it (qubit 1).
		\item Lift-off in a NEP bath ($1\,\hour$ at $90\,^\circ\mathrm{C}$).
	\end{enumerate}	
\item dicing as described in 1 h.
\end{enumerate}

\begin{table}[hp]
\centering
\begin{tabular}{|c|c|}
\hline
Device & Model \\ \hline\hline 
electron beam writer & JEOL JBX-5500ZD\\
  & $50\,\mathrm{keV}$ acceleration voltage\\ \hline
mask aligner &  Carl Suess MA6, Xe $500\,\mathrm{W}$ lamp,\\
			& wave lengths: $240,\;365,\;405\,\mathrm{nm}$\\ \hline
inductively coupled & Oxford Plasma Technology\\
 plasma (ICP) device & Plasmalab 100 ICP180\\
 &variable substrate temp. $273-343\,\mathrm{K}$\\
  \hline 
shadow evaporation device & PLASSYS MEB550s\\
    \hline
O$_2$ cleaner ("Barrel asher") & "nano", Diener electronic GmbH\\
&$40\,\kilo\mathrm{Hz}$ generator, $0-300\,\watt$\\
    \hline  
\end{tabular}
\caption{Apparatus used at KIT to fabricate qubit samples.}
\label{tab_fab}
\end{table}

\begin{table}[h!]
\centering
\begin{tabular}{|c|c|c|c|c|}
\hline
\multicolumn{5}{|l|}{\textbf{Descum} ($10\,\mathrm{sccm}$ O$_2$, $5\,\mathrm{sccm}$ Ar)}\\ \hline
Cathode & Discharge & Beam & acceleration & neutralizer \\
$8\,\volt$ & $40\,\volt$ & $50\,\volt$ & $11\,\volt$ & $5\,\volt$ (emission) \\
$5\,\ampere$ & $0.1\,\ampere$ & $5\,\milli\ampere$ & $4\,\milli\ampere$ & $9\,\volt$ \\ 
& & & & $11\,\milli\ampere$ \\ \hline
\end{tabular}
\caption{Parameters of the KSC 1202 power supply unit that controls the KDC 40 ion source to generate the remote Oxygen-Argon plasma used to descum or clean samples in the PLASSYS MEB550s shadow evaporation tool.}
\label{tab_descum}
\end{table}

\begin{table}[h!]
\centering
\begin{tabular}{|c|c|c|c|c|}
\hline
\multicolumn{5}{|l|}{\textbf{Argon ion-milling} ($4\,\mathrm{sccm}$ Arr)}\\ \hline
Cathode & Discharge & Beam & acceleration & neutralizer \\
$7\,\volt$ & $50\,\volt$ & $400\,\volt$ & $91\,\volt$ & $15\,\volt$ (emission) \\
$4\,\ampere$ & $0.1\,\ampere$ & $15\,\milli\ampere$ & $0.9\,\milli\ampere$ & $9\,\volt$ \\ 
& & & & $1\,\milli\ampere$ \\ \hline
\end{tabular}
\caption{Parameters of the KSC 1202 power supply unit that controls the KDC 40 ion source to generate the Argon ion-milling remote plasma in the PLASSYS MEB550s shadow evaporation device.}
\label{tab_armill}
\end{table}

\section*{Supplementary Methods II}
\label{sec:densities}
\noindent\textbf{Measurement of defect densities}\\
The spectral distribution of defects has been recorded using the swap spectroscopy protocol (see inset of Fig.~2\,\textbf{b}): the qubit is excited with a pi pulse, and tuned to a given frequency for a constant duration. If the qubit is near resonance with a defect, an enhanced energy relaxation is detected since the defects detectable in superconducting qubits have typically very short coherence times ($\sim100\,\nano\second$)~\cite{Lisenfeld19,Barends13}. This routine is repeated several hundreds times for statistics, and for varying frequencies to obtain the spectral distribution of defects, as shown in Figure~2 of the main text.\\

\begin{figure}[htbp]
		\includegraphics[width=\linewidth]{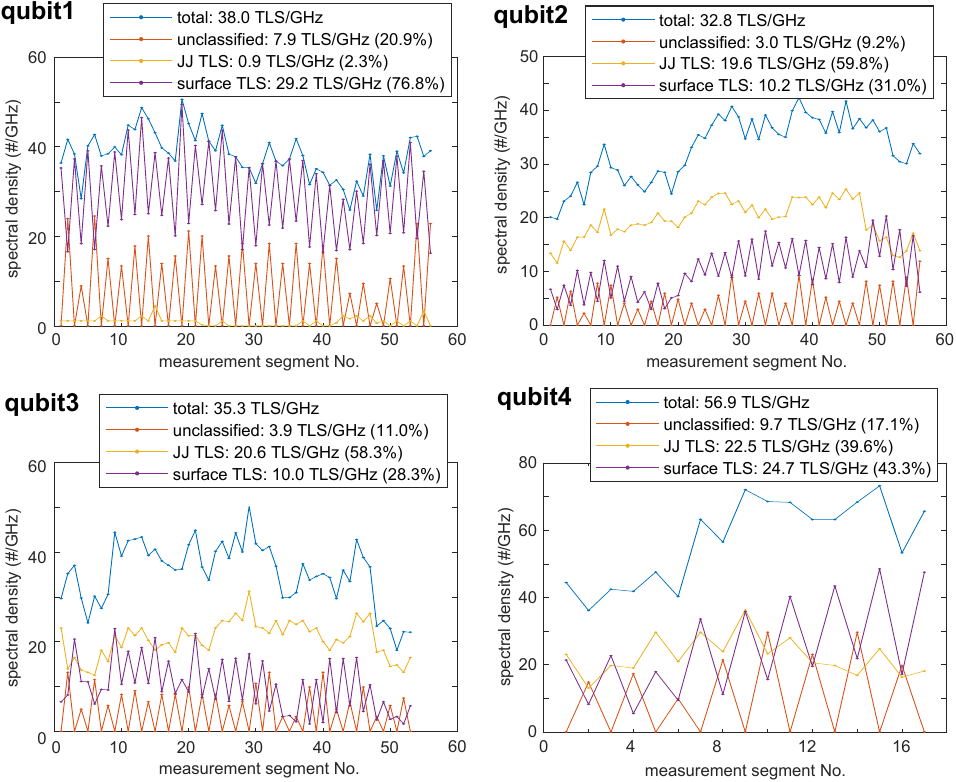}
		\caption{Mean spectral density of defects per measurement segment of the defect spectroscopy data shown in Fig. 2 of the  main text. The legend contains the averaged  spectral density and the relative portion of defects residing in junctions, at surfaces, as well as unclassified ones.
}
	\label{fig:S5}
\end{figure}

Figure~\ref{fig:S5} contains the average spectral defect densities of chip \#1 in each measurement segment while the legends show the mean values and relative contributions. The junction-defects are identified by their zero response to the global gate (control voltage \vg). In turn, all defects are sensitive to strain (control voltage \vp). Thus, a defect's location cannot be identified if it is detectable only in one measurement segment vs. the piezo voltage. This is gives the spectral density of unclassified defects ("dead counts") which is zero in data segments where \vg was swept, and non-zero in segments which were recorded for swept \vp. This explains the oscillations visible in all panels of Fig.~\ref{fig:S5}. The amount of dead counts increases with the segment width (see qubit No 4) since the probability for a single defect to be tuned out of the qubit's tunability range is higher for wider segments. On the other side, choosing too narrow segments again increases the number of dead counts due increasing complexity of sorting out individual defect curves from the data. For instance, in the data set recorded with qubit 1, the segments were narrow, and the surface-defect density high, which explains the relatively high amount of dead counts. The segment width in the data set of qubit 4 represents a good trade-off between analysis complexity and dead counts.\\

\begin{figure}[htbp]
		\includegraphics[width=\linewidth]{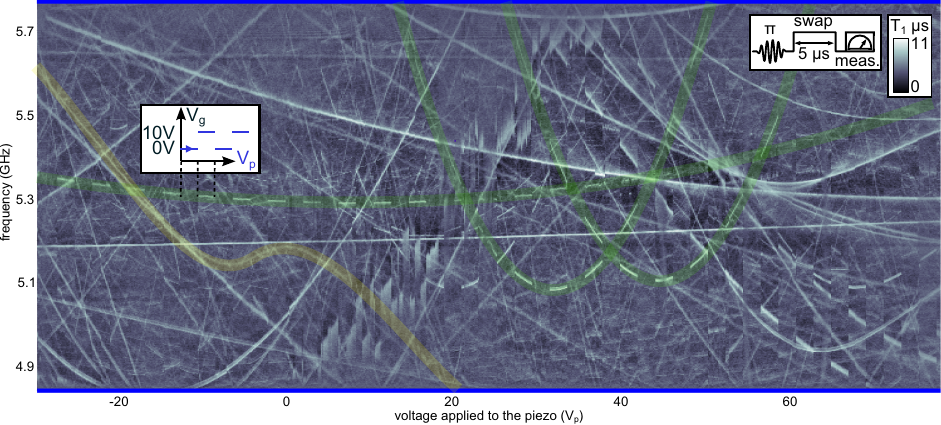}
		\caption{Defect spectroscopy recorded with qubit 2 of sample \#2 with a method highlighting defects in Josephson junctions.	As explained in the text, the traces of junction-defects appear as continuous lines while non-junction defects appear as dashed lines (see some exemplary green lines). The trace highlighted in yellow indicates a rare case of mutually interacting TLS as described in ~\cite{Lisenfeld2015}.
}
	\label{fig:S6}
\end{figure}

The defect densities of sample \#2 were obtained using a similar method described above, which however was optimized to identify junction-defects. This procedure results in data as shown in Fig.~\ref{fig:S6}, obtained on qubit 2.2 (sample \#2, qubit 2). The swap spectroscopy protocol (see right inset) was repeated at various values $V_\text{p}$ of the voltage applied to the piezo actuator. Each time after $V_\text{p}$ was increased by 10V,  $V_\text{g}$ ($\propto$ the applied electric field) was switched to either 0 or 10V in an alternating manner as indicated in the inset. Since junction-defects are insensitive to the gate, their traces appear as continuous lines, while non-junction defects appear as dashed lines, as highlighted in green color. The color bar encodes an estimated value for the qubit's $T_1$ time. In this data set we see a defect which is apparently z-coupled to another defect~\cite{Lisenfeld2015} that is not detectable by the qubit (yellow line).\\

\clearpage
\section*{Supplementary Discussion I}
\label{ss_defectdensities_surf}

\begin{figure}[htbp]
		\includegraphics[width=.95\linewidth]{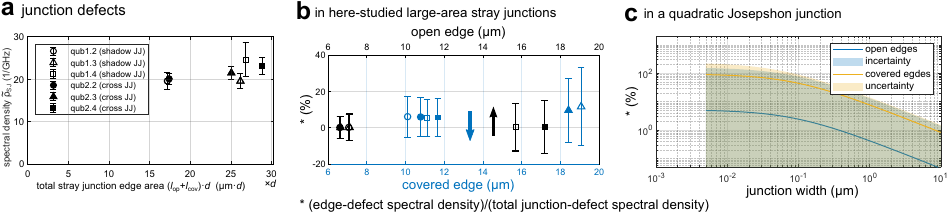}
		\caption{\textbf{a} Measured spectral density of junction-defects vs. the stray junction's total edge area. \red{\textbf{b} Relative share of junction-defects in here-studied stray junctions. \textbf{c} estimated relative share of edge-defects in quadratic small junctions, based on the fit results presented in Eq. (2) in the main text. The purpose of this figure is merely to indicate that edge-defects may dominate the number of defects in sub-micron junctions, however note the huge uncertainty due to the fit error quoted in Eq. (2). Possibly, a larger data set could improve the fit error.}
	}
	\label{fig:S2}
\end{figure}

\noindent{\textbf{Junction-defects}}\\
The plot in Fig.~\ref{fig:S2} \textbf{a} shows the density of detectable junction-defects as a function of the total area of the junction edges (total edge length times the effective edge width $d\approx2\,\mathrm{nm}$). A fit of the merged data set from both samples to $\rhos = \rhoa A + \rho_\text{totedge} (l_\text{op}+l_\text{cov})d$ in dependence of the junction area and total edge area returns $\rhoa=1.5\pm0.3\,\mathrm{(GHz\cdot\mu m^2)^{-1}}$ and $\rho_\text{totedge}=32.3\pm92.1\,\mathrm{(GHz\cdot\mu m^2)^{-1}}$. The large fit error to $\rho_\text{totedge}$ indicates, similar to the result in Eq. (2) in the main text that the observed defect spectral density does not significantly scale with the junction edge area. As a note, it seems that the data in Fig.~\ref{fig:S2} \textbf{a} fits to a linear function with some offset which however is not physical as verified with the reference qubits 1.1 and 2.1 which have very small junctions and show a very small density of detectable junction-defects.\\

As a note to Fig. 3~\textbf{a} in the main text, the linear plot $\rhosj = \rhoa A$ (orange line) seems to not ideally fit the orange data points. This is not surprising since the coefficients $\rhoa$, $\rhoo$ and $\rhoc$ stem from a linear fit to the four dimensional expression shown in Eq.~(2), whereas the orange plot is a projection on a two-dimensional plane.\\

\red{Figure~\ref{fig:S2} \textbf{b} contains the relative share of edge-defects in the here-studied large-area stray Josephson junctions. The average value at covered and open junction edges are quoted in the main text. The high uncertainty is due to the high fit error presented in Eq. (2) of the main text. Based on the fit results shown in Eq. (2) we have extrapolated the relative share of edge-defects in smaller and quadratic junctions, as shown in Fig.~\ref{fig:S2} \textbf{c}. This estimation indicates that for junctions sized smaller than $(100\,\mathrm{nm})^2$ the edges could predominantly contribute to the amount of junction-defects (i.e. the relative share of junction-defects approaches $100\,\%$), which is however uncertain due to a huge error. We thus conclude that based on our data set we cannot make reliable predictions whether or not junction edges dominate dielectric losses in small junctions.}\\

\section*{Supplementary Discussion II}
\label{ss_defectdensities_JJ}

\begin{figure}[htbp]
		\includegraphics[width=.8\linewidth]{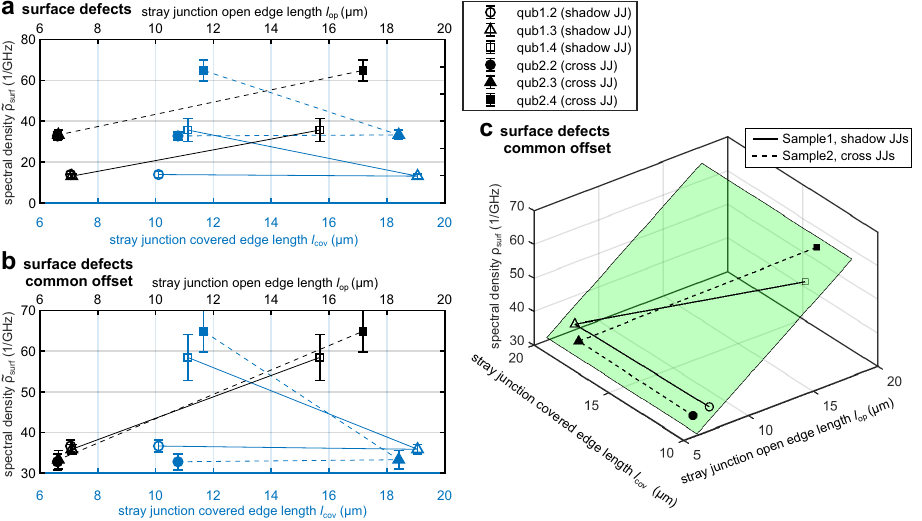}
		\caption{\textbf{a} Spectral density of non-junction defects (defects residing outside the Josephson junction) plotted vs. the length of the open (black) and the covered (blue) edge of the stray junction. \red{The straight lines are guides for the eye (continuous: sample \#1, dashed: sample \#2). \textbf{b} Same data sets from sample \#1 and \#2, which are brought to a common offset. \textbf{c} Data from \textbf{b} in a three-dimensional plot, where the green plane represents the fit results to Eq.~\eqref{eq_S1}. A linear dependence is only observed versus the open junction edge.}
	}
	\label{fig:S1}
\end{figure}

\noindent{\textbf{Surface-defects}}\\
Fig.~\ref{fig:S1} \textbf{a} contains a plot of the measured spectral density of surface-defects (all defects which reside outside of the Josephson junctions, i.e. non-junction defects) vs. the open and covered stray junction edge length, where the data points are interconnected with continuous lines to guide the eye, \red{ and where the error bar $\rhonc\rhosurf/(\rhosj+\rhosurf)$ is the spectral density of non-classified detected defects \rhonc~multiplied with the relative part of surface-defects.} We see that the number of detected surface-defects predominantly scales with the length of the open junction edge. This is further supported by the linear fit of the expected surface-defect density
\begin{align}
\rhosurf=\rhoos l_\text{op} + \rhocs l_\text{cov} + const \label{eq_S1}
\end{align}
to the plotted data points, which results in $\rhoos=(2.54\pm0.03)\,\mathrm{(GHz\cdot\mu m)^{-1}}$ and $\rhocs=(-0.10\pm0.18)\,\mathrm{(GHz\cdot\mu m)^{-1}}$ for the sample \#1, and results in $\rhoos=(3.03\pm0.17)\,\mathrm{(GHz\cdot\mu m)^{-1}}$ and $\rhocs=(0.05\pm0.12)\,\mathrm{(GHz\cdot\mu m)^{-1}}$ for data from the sample \#2. \red{ The constant in Eq.~\eqref{eq_S1} is due to surface-defects on the qubit electrodes. We see in Fig.~\ref{fig:S1} \textbf{a} that both data sets show similar proportionality versus the junction edges, so we bring them to a common offset as shown in Fig.~\ref{fig:S1} \textbf{b}, in order to improve the linear fit quality to Eq.~\eqref{eq_S1}. The fit proposes similar results $\rhoos=(2.86\pm0.2)\,\mathrm{(GHz\cdot\mu m)^{-1}}$ and $\rhocs=(0.03\pm0.18)\,\mathrm{(GHz\cdot\mu m)^{-1}}$(offset not important), which are represented by a green plane in Fig.~\ref{fig:S1}~\textbf{c}.}\\

\end{document}